\documentclass[preprint]{aastex}

\shorttitle{Constraints on the Heating of High Temperature Loops}
\shortauthors{Warren et al.}

\begin{document}


\title{Constraints on the Heating of High Temperature Active Region Loops: \\
  Observations from \textit{Hinode} and \textit{SDO}}

\author{Harry P. Warren\altaffilmark{1}, David H. Brooks\altaffilmark{2}, and Amy
  R. Winebarger\altaffilmark{3}}

\altaffiltext{1}{Space Science Division, Naval Research Laboratory, Washington, DC 20375}
\altaffiltext{2}{College of Science, George Mason University, 4400 University Drive,
  Fairfax, VA 22030}
\altaffiltext{3}{NASA Marshall Space Flight Center, VP 62, Huntsville, AL 35812}


\begin{abstract}
  We present observations of high temperature emission in the core of a solar active
  region using instruments on \textit{Hinode} and \textit{SDO}.  These multi-instrument
  observations allow us to determine the distribution of plasma temperatures and follow
  the evolution of emission at different temperatures.  We find that at the apex of the
  high temperature loops the emission measure distribution is strongly peaked near 4\,MK
  and falls off sharply at both higher and lower temperatures. Perhaps most significantly,
  the emission measure at 0.5\,MK is reduced by more than two orders of magnitude from the
  peak at 4\,MK. We also find that the temporal evolution in broad-band soft X-ray images
  is relatively constant over about 6 hours of observing. Observations in the cooler
  \textit{SDO}/AIA bandpasses generally do not show cooling loops in the core of the
  active region, consistent with the steady emission observed at high temperatures. These
  observations suggest that the high temperature loops observed in the core of an active
  region are close to equilibrium. We find that it is possible to reproduce the relative
  intensities of high temperature emission lines with a simple, high-frequency heating
  scenario where heating events occur on time scales much less than a cooling time. In
  contrast, low-frequency heating scenarios, which are commonly invoked to describe
  nanoflare models of coronal heating, do not reproduce the relative intensities of high
  temperature emission lines and predict low-temperature emission that is approximately an
  order of magnitude too large.  We also present an initial look at images from the
  \textit{SDO}/AIA 94\,\AA\ channel, which is sensitive to \ion{Fe}{18}.
\end{abstract}



  \section{Introduction}

  Understanding the origin of high temperature plasma in the solar corona is one of the
  central problems in solar physics. The nanoflare concept represents one of the more
  popular theories for describing how energy stored in the Sun's magnetic field is
  converted into thermal energy \cite[e.g.,][]{parker1972,parker1983}. In nanoflare models
  turbulent photospheric motions drive the twisting and braiding of the magnetic field,
  which leads to the release of energy on small spatial scales as the accumulated
  topological complexity is dissipated by magnetic reconnection. Nanoflares are often
  modeled as impulsive heating events where the plasma is heated to very high temperatures
  ($\sim10$\,MK) and then cools down through a combination of conduction, enthalpy flux,
  and radiation without being reheated. Furthermore, since these heating events are likely
  to occur on very small spatial scales an observed loop is assumed to be composed of many
  unresolved strands \cite[e.g.,][]{cargill1994,klimchuk2001,patsourakos2006}. This
  framework implies that nanoflare heated loops should have co-spatial hot and cool
  emission.

  Observations at relatively cool coronal temperatures, however, have have cast doubt on
  this heating scenario. \cite{antiochos2003} and \cite{nitta2000}, for example, have
  argued that cooling loops are often not observed in the core of an active region. This
  position has been supported by recent observations with the EUV Imaging Spectrograph
  (EIS) and the X-ray Telescope (XRT) on the \textit{Hinode} mission
  \citep{kosugi2007}. \cite{warren2010} found no evidence for emission near 1\,MK that was
  spatially correlated with emission at higher temperatures (3--5\,MK) in the core of an
  active region. In this region the high temperature emission measured with XRT was
  observed to be relatively constant over many hours. Similarly, \cite{brooks2009} have
  found no evidence for strong variability in the intensities and Doppler shifts measured
  in the moss, the footpoints of the high temperature loops. These results suggest that
  high temperature loops are generally heated on very short time scales, much shorter than
  a characteristic cooling time, and do not cool down to lower temperatures.

  \begin{figure*}[t!]
  \centerline{\includegraphics[clip,scale=1]{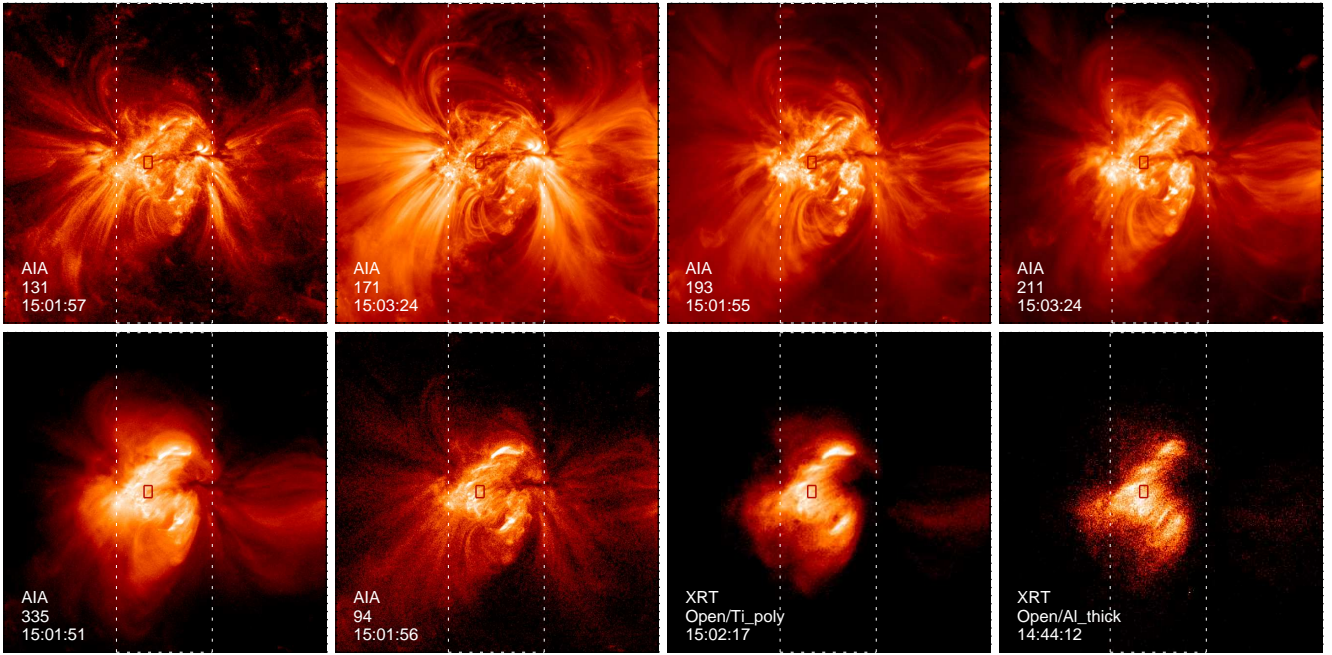}}
  \caption{AIA and XRT observations of active region 11089 taken on 2010 July 23 near 15
    UT. The field of view is $384\arcsec\times384\arcsec$, which is motivated by the size
    of the XRT images. The dotted line indicates the extent of the EIS raster. The small
    box indicates the region between the moss used to calculate the emission measure with
    EIS and XRT, and represents the physical conditions at the loop apexes along the line
    of sight. This region was chosen because it minimizes contamination from the moss.
    The electronic version of the manuscript includes movies of the AIA 171, 335, 94, and
    131\,\AA\ channels.}
  \label{fig:aia}
  \end{figure*}

  Previous studies on the properties of high temperature emission in the core of an active
  region have failed to adequately address a fundamental question: what is the
  distribution of temperatures in the core of an active region? The EIS spectral range
  contains emission lines from \ion{Ca}{14}--\ion{Ca}{17} which provide excellent coverage
  of the critical 3--5\,MK temperature range (see
  \citealt{delzanna2008,warren2008}). Observations of these lines in combination with
  other emission lines observed with EIS and observations with the thick XRT filters allow
  for the emission measure distribution to be computed over a very wide range of
  temperatures. The properties of the loop apexes along the line of sight are of
  particular importance. The dispersion in the temperature constrains the rate at which
  energy is input into high temperature loops and such information is critical for
  motivating physical models of this emission.

  In this paper we present the analysis of new observations in the core of an active
  region taken after the launch of the \textit{Solar Dynamics Observatory}
  (\textit{SDO}). These observations feature spectroscopic observations from EIS and soft
  X-ray images from XRT on \textit{Hinode}. These data show that the emission measure
  distribution for the loop apex is peaked at temperatures near 4\,MK and falls off
  sharply at both higher and lower temperatures. At temperatures near 0.5\,MK the emission
  measure is more than two orders of magnitude below the value at 4\,MK.  Simple
  hydrodynamic modeling suggests that these properties are inconsistent with low frequency
  nanoflare heating models. Such models predict low temperature emission that is
  approximately an order of magnitude larger than what is observed.

  The high cadence ($\sim12$\,s), high spatial resolution ($0.6\arcsec$ pixels), narrow
  band images provided by the Atmospheric Imaging Assembly (AIA) on \textit{SDO} provide
  additional information on the evolution of plasma at a wide range of
  temperatures. Images from channels dominated by 1\,MK emission show only a few, isolated
  cooling loops in the moss regions, consistent with previous results. In this paper we
  also present some initial observations from the AIA 94\,\AA\ channel, which is sensitive
  to the \ion{Fe}{18} 93.94\,\AA\ line formed at about 7\,MK. In small flares and
  microflares the expected progression from high temperature to low temperature emission
  is observed. In more quiescent areas of the active region core, however, the emission
  appears to be relatively constant, potentially consistent with what is observed in XRT.

 \section{Observations and Analysis}

  \begin{figure*}[t!]
  \centerline{\includegraphics[clip,scale=1]{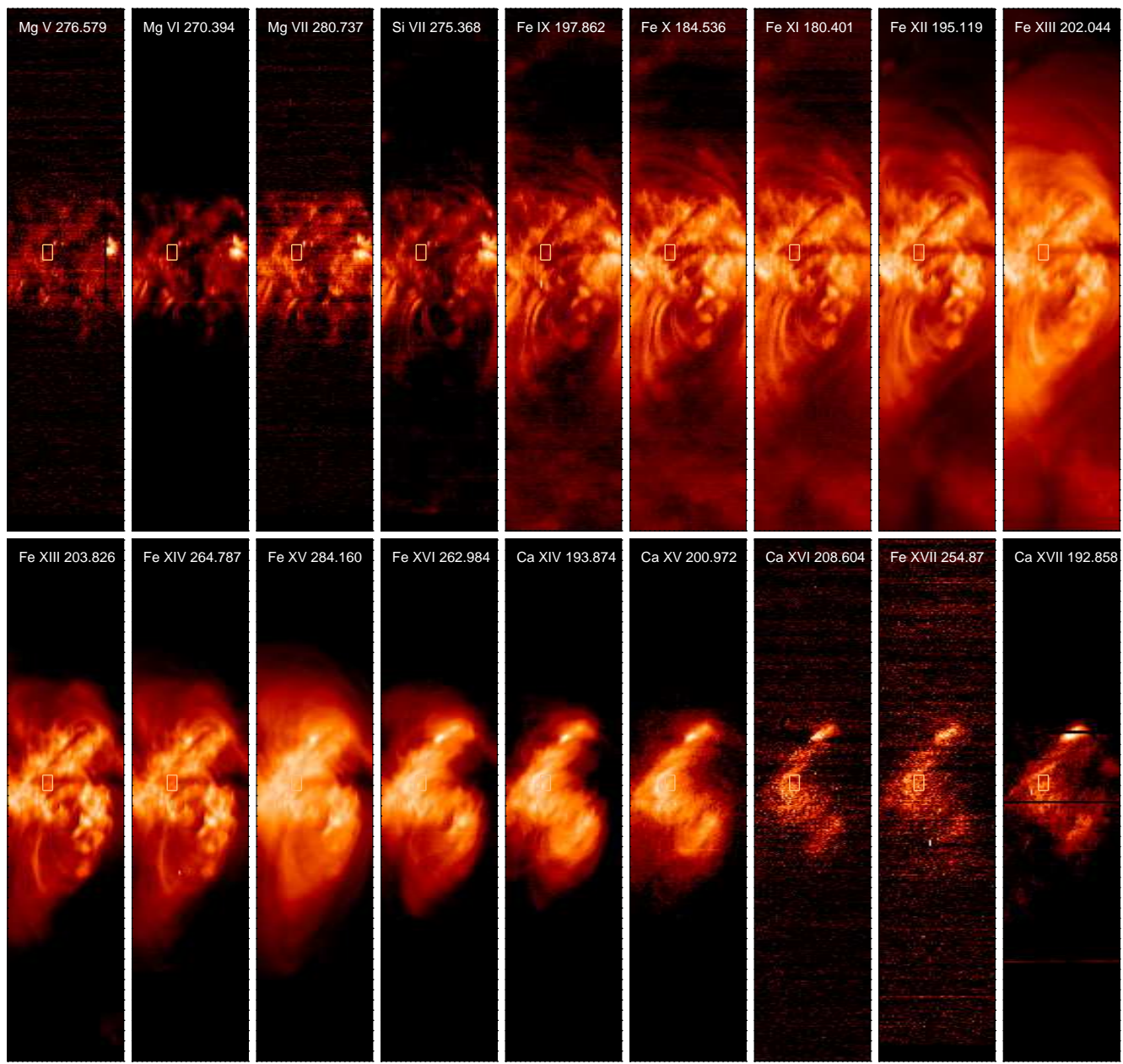}}
  \caption{EIS observations of AR 11089 in various emission lines. The field of view is
    $120\arcsec\times384\arcsec$. The small box indicates the region between the moss used
    to calculate the active region ``core'' emission measure. The size of this region is
    $10\arcsec\times15\arcsec$ or 75 spatial pixels.}
  \label{fig:eis}
  \end{figure*}

  \begin{figure*}[t!]
  \centerline{\includegraphics[clip,scale=0.5]{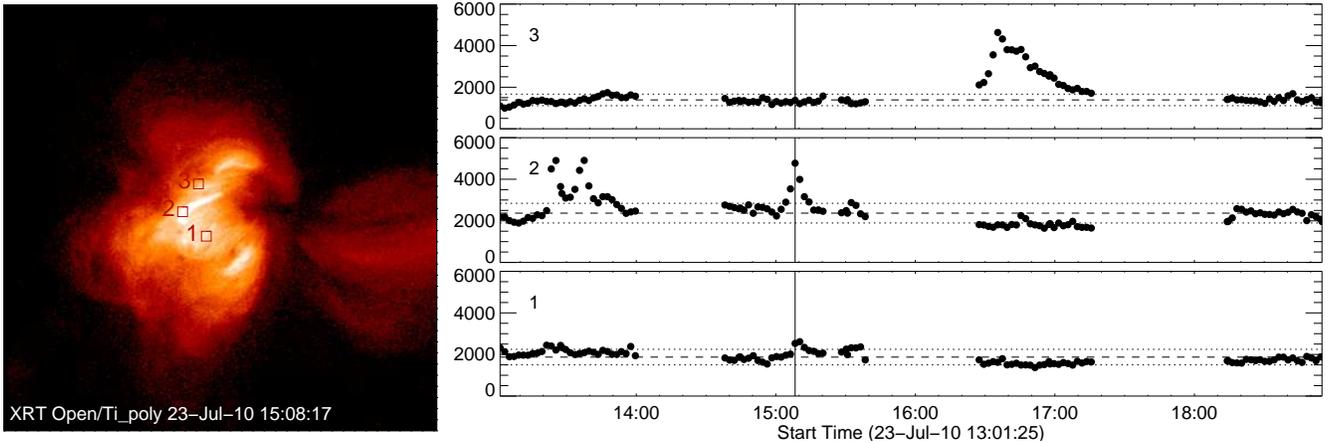}}
  \caption{XRT observations of AR 11089. On the left is a frame from the observing
    sequence. On the right are light curves from three different locations in the core of
    the active region. Intensities are in DN~s$^{-1}$ per pixel. The dashed horizontal
    line is the median intensity for the period shown. The dotted horizontal lines are
    $\pm20\%$ of the median. The solid vertical line corresponds to the time of the
    image. The emission in the core of the active region is relatively constant, with a
    variability of 20\% or less. Transient events with relatively short lifetimes
    ($\lesssim900$\,s) are also observed. The electronic version of the manuscript
    includes a movie of these data.}
  \label{fig:xrt}
  \end{figure*}
  
  In this paper we consider EIS, XRT, and AIA observations of NOAA active region 11089
  taken on 2010 July 23. This section gives a brief overview of the various instruments,
  the available data, and the calculation of the differential emission measure
  distribution.

  The AIA on \textit{SDO} consists of 4 independent normal incidence, multi-layer
  telescopes, each with 2 channels. Each channel images the full Sun. The AIA coronal
  channels and their dominant components are as follows: 94\,\AA\ (\ion{Fe}{10} and
  \ion{Fe}{18}), 131\,\AA\ (\ion{Fe}{8} and \ion{Fe}{20}, \ion{Fe}{21}, and \ion{Fe}{23}),
  171\,\AA\ (\ion{Fe}{9} and \ion{Fe}{10}), 193\,\AA\ (\ion{Fe}{12}), 211\,\AA\
  (\ion{Fe}{14}), 335\,\AA\ (\ion{Fe}{16}). A detailed discussion of the predicted
  contributions to each AIA channel is presented by \cite{odwyer2010}. The relatively cool
  components of the 94 and 131\,\AA\ channels have been identified in high spectral
  resolution stellar observations \citep[e.g.,][]{raassen2002,liang2010}. There is also a
  \ion{He}{2} 304\,\AA\ channel and a UV channel. Figure~\ref{fig:aia} shows sample AIA
  images from this active region region taken near 15 UT. For this work we use the level
  1.5 cutout data that has been re-scaled and co-aligned to a common coordinate system. We
  note that to precisely co-align the data in the various channels we cross-correlated a
  subset of 20 full-disk images taken from the period of interest. These offsets were
  generally on the order of 1--2 pixels in each direction. De-rotation of the data was
  achieved by using established solar rotation rates. As is evident in the movies, there
  is very little jitter in the data.

  The XRT on \textit{Hinode} \citep{golub2007} is a grazing incidence, soft X-ray
  telescope. Temperature discrimination is achieved through the use of focal plane
  filters. Because XRT can observe the Sun at short wavelengths, XRT can observe high
  temperature solar plasma very efficiently. For the observations considered here, XRT
  took images in the thin Ti-poly filter at a cadence of about 120\,s, with Al-thick
  context images taken approximately every 90 minutes. The observing was periodically
  interrupted by seasonal eclipses that occur annually during the May -- August
  period. Representative XRT Ti-poly and Al-thick images are shown in
  Figure~\ref{fig:aia}. The XRT images have been co-aligned with respect to the first
  image in the sequence. The XRT and AIA data were co-aligned by cross-correlating the
  Ti-poly and 335\,\AA\ images. This analysis incorporates the latest available
  calibration for XRT \citep{narukage2010}.

  The EIS instrument on \textit{Hinode} \citep{culhane2007,korendyke2006} is a high
  spatial and spectral resolution imaging spectrograph. EIS observes two wavelength
  ranges, 171--212\,\AA\ and 245--291\,\AA, with a spectral resolution of about 22\,m\AA\
  and a spatial resolution of about 1\arcsec\ per pixel.  Solar images can be made by
  stepping the slit over a region of the Sun. Telemetry constraints generally limit the
  spatial and spectral coverage of an observation. For the observations considered here
  the 1\arcsec\ slit was stepped over the core of the active region using 60 2\arcsec\
  steps and 512\arcsec\ of the slit height was read out. The field of view for the EIS
  raster that was run between 14:32 and 15:34 UT on 2010 July 23 is indicated in
  Figure~\ref{fig:aia}. 

  The use of wide 2\arcsec\ steps in the EIS raster degrades the spatial resolution
  somewhat, but allows for the raster to be completed in between \textit{Hinode} eclipses
  even while using relatively long (60\,s) exposure time. The use of wide steps also
  reduces telemetry usage and allows for a very extensive line list to be telemetered to
  the ground. From these data we have determined intensities for 39 spectral lines. Images
  for some of these emission lines are shown in Figure~\ref{fig:eis}. Almost all of the
  line intensities can be determined by simple Gaussian fits. The \ion{Ca}{17}
  192.858\,\AA\ line is complicated by blending with \ion{Fe}{11} 192.813\,\AA\ and a
  complex of \ion{O}{5} lines, including \ion{O}{5} 192.906\,\AA. The intensities for
  these lines are determined using the method outlined by \cite{ko2009}, where the
  intensity of the \ion{Fe}{11} 192.813\,\AA\ line is inferred from \ion{Fe}{11}
  188.219\,\AA\ and the remaining part of the profile is fit with multiple Gaussians. The
  resulting \ion{Ca}{17} raster is consistent with \ion{Fe}{17} 254.87\,\AA, indicating
  that the deconvolution is approximately correct.

  Collectively, these data suggest a relatively simple morphology to the active
  region. The EIS emission lines formed at high temperatures, \ion{Fe}{16} (2.82\,MK),
  \ion{Ca}{14} (3.55\,MK), \ion{Ca}{15} (4.47\,MK), \ion{Ca}{16} (5.01\,MK), and
  \ion{Ca}{17} (5.62\,MK), and the XRT Ti-poly and Al-thick images show relatively short,
  hot loops that connect across the neutral line running through the middle of the active
  region. The temperatures given here correspond to the peaks in the CHIANTI 6.0
  ionization fractions \citep{dere2009}. Below these hot loops are regions of intense
  ``moss,'' the footpoints of the hot loops that are evident in emission lines formed at
  temperatures near 1\,MK
  \citep[e.g.,][]{peres1994,berger1999,fletcher1999,martens2000}. As mentioned earlier,
  the relatively cool fan-like loops \citep[e.g.,][]{schrijver1999} appear to be largely
  absent from the core of the active region. The emission at these temperatures, however,
  is not zero. There is certainly diffuse million degree coronal emission lying above the
  active region. The AIA movies also indicate that there is some million degree emission
  associated with the very cool, and highly dynamic filament-like material in the core of
  the active region that connects across the neutral line at very low heights.

  As mentioned previously, the primary strength of the XRT is the ability to observe high
  temperature emission over a wide field of view at relatively high cadence. This is
  illustrated in Figure~\ref{fig:xrt}, where light curves from several positions in the
  core of the active region are shown. Each light curve is from a single pixel, there is
  no spatial averaging. Consistent with earlier analysis \cite[e.g.,][]{warren2010}, these
  light curves show relatively constant emission. Typical intensities generally lie within
  a range of $\pm20\%$ over many hours. Transient brightenings are observed, but they are
  relatively short lived, with lifetimes of 900\,s or less.

  The AIA movies, which are included in the electronic version of this manuscript, show
  that the moss is consistent with the relatively constant emission observed in XRT.  The
  AIA 131 and 171\,\AA\ movies show clear evidence for many fine coronal loops in the
  active region. Few of these loops, however, connect to elements of the moss. There is
  evidence for intensity fluctuations in the moss itself, but it is likely that this is
  related to spicular activity that obscures the moss \citep{depontieu1999}.
  
  The primary objective of this study is to determine the distribution of temperatures in
  the hot loops in the core of the active region. We are particularly interested in the
  conditions at the loop apexes along the line of sight that avoids contamination from the
  moss emission. Such measurements are important for determining the physical conditions
  in the high temperature loops. To accomplish this we have determined spatially averaged
  EIS line profiles in the small ``inter-moss'' region indicated in Figures~\ref{fig:aia}
  and \ref{fig:eis}. The size of this region is $10\arcsec\times15\arcsec$ or 75 spatial
  pixels when we account for the 2\arcsec\ steps. This region has relatively little
  contamination from the moss or from overlying loops. The spatial averaging diminishes
  the uncertainties for the intensities of some of the weaker lines and allows for a more
  complete analysis. The intensities for these line profiles are computed using the same
  assumptions as the spatially resolved intensities shown in the rasters. The intensities
  from this region are given in Table~\ref{table:ints1}. The selection of lines is
  motivated by previous emission measure calculations in the quiet Sun
  \citep{brooks2009b,warren2009} and in flares \citep{delzanna2008,warren2008}. This
  previous work has identified emission lines that are mutually consistent. Note that
  \ion{Fe}{17} 254.87\,\AA\ has been left out because of uncertainties in the atomic data
  \citep{warren2008}.

  \begin{figure*}[t!]
  \centerline{\includegraphics[clip,scale=1]{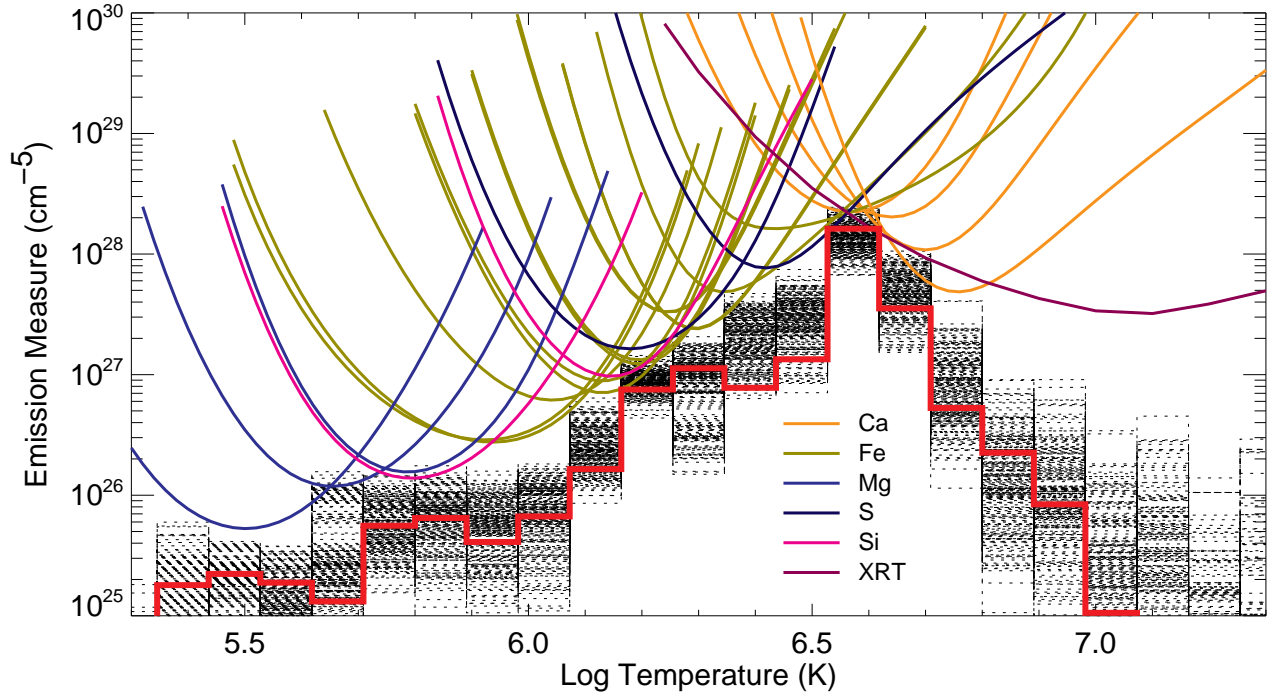}}
  \caption{The emission measure distribution for the core of the active region. The DEM
    was computed using the ``inter-moss'' intensities from EIS and XRT and represents the
    physical conditions at the loop-tops along the line of sight. The solid red line is
    the best-fit DEM and the dotted black lines are from Monte Carlo calculations using
    perturbed values of the observed intensities. 250 Monte Carlo simulations of the
    emission measure were performed. This gives a statistically plausible range for the
    emission measure in each temperature bin. The solid lines are the emission measure
    loci curves described in the text and color-coded according to element. The emission
    measure distribution is strongly peaked at $\log T = 6.6$ and falls off sharply for
    both higher and lower temperatures.}
  \label{fig:dem}
  \end{figure*}

\begin{deluxetable}{lcrrrr}
\tablewidth{0in}
\tabletypesize{\small}
\tablecaption{Differential Emission Measure Model\tablenotemark{a}}
\tablehead{
   \multicolumn{1}{c}{Line}      &
   \multicolumn{1}{c}{$T_{max}$} &
   \multicolumn{1}{c}{$I_{obs}$} &
   \multicolumn{1}{c}{$\sigma_{I}$} &
   \multicolumn{1}{c}{$I_{dem}$} &
   \multicolumn{1}{c}{$R$}
}
\startdata
   \ion{Mg}{5} 276.579 &       5.45 &       16.5 &        3.8 &       19.5 &       0.85 \\
   \ion{Mg}{6} 270.394 &       5.65 &       35.9 &        7.9 &       27.6 &       1.30 \\
   \ion{Mg}{7} 280.737 &       5.80 &       32.7 &        7.4 &       29.7 &       1.10 \\
   \ion{Si}{7} 275.368 &       5.80 &       47.0 &       10.4 &       54.3 &       0.87 \\
   \ion{Fe}{9} 197.862 &       5.85 &       40.0 &        8.8 &       39.7 &       1.01 \\
   \ion{Fe}{9} 188.497 &       5.85 &       72.4 &       16.0 &       67.0 &       1.08 \\
  \ion{Fe}{10} 184.536 &       6.05 &      280.4 &       61.8 &      202.0 &       1.39 \\
  \ion{Si}{10} 258.375 &       6.15 &      294.0 &       64.7 &      321.2 &       0.92 \\
  \ion{Fe}{11} 188.216 &       6.15 &      578.2 &      127.2 &      556.0 &       1.04 \\
  \ion{Fe}{11} 180.401 &       6.15 &      926.1 &      204.2 &     1120.2 &       0.83 \\
   \ion{S}{10} 264.233 &       6.15 &       71.8 &       15.9 &       73.9 &       0.97 \\
  \ion{Fe}{12} 195.119 &       6.20 &     1475.4 &      324.6 &     1491.6 &       0.99 \\
  \ion{Fe}{12} 192.394 &       6.20 &      437.8 &       96.3 &      478.5 &       0.91 \\
  \ion{Fe}{13} 202.044 &       6.25 &     1248.3 &      274.7 &      665.9 &       1.87 \\
  \ion{Fe}{13} 203.826 &       6.25 &     2533.9 &      557.6 &     1331.4 &       1.90 \\
  \ion{Fe}{14} 270.519 &       6.30 &      515.0 &      113.3 &      523.0 &       0.98 \\
  \ion{Fe}{14} 264.787 &       6.30 &     1026.9 &      226.0 &     1028.8 &       1.00 \\
  \ion{Fe}{15} 284.160 &       6.35 &    10334.0 &     2273.6 &    11634.3 &       0.89 \\
   \ion{S}{13} 256.686 &       6.40 &      854.7 &      188.1 &      853.2 &       1.00 \\
  \ion{Fe}{16} 262.984 &       6.45 &     1157.6 &      254.7 &     1138.9 &       1.02 \\
  \ion{Ca}{14} 193.874 &       6.55 &      311.9 &       68.6 &      276.3 &       1.13 \\
  \ion{Ca}{15} 200.972 &       6.65 &      238.9 &       52.6 &      195.7 &       1.22 \\
  \ion{Ca}{16} 208.604 &       6.70 &      122.0 &       28.3 &      108.7 &       1.12 \\
  \ion{Ca}{17} 192.858 &       6.75 &      146.5 &       32.3 &      128.5 &       1.14 \\
         Open/Al-thick &       7.10 &        4.5 &        0.9 &        5.8 &       0.78
\enddata
\tablenotetext{a}{Calculated intensities are from the MCMC emission measure
  inversion. Intensities are in units of erg cm$^{-2}$ s$^{-1}$ sr$^{-1}$.}
\label{table:ints1}
\end{deluxetable}

  To compute the differential emission measure we use the Monte Carlo Markov Chain (MCMC)
  emission measure algorithm \citep{kashyap1998,kashyap2000} distributed with the
  \verb+PINTofALE+ spectral analysis package. This algorithm has the advantage of not
  assuming a functional form for the differential emission measure. The MCMC algorithm
  also provides for estimates of the error in the DEM by calculating the emission measure
  using perturbed values for the intensities. The algorithm assumes the uncertainties in
  the intensities are uncorrelated so that systematic errors in the calibration, which
  could depend on the wavelength, or in the atomic data, which could vary by ion, are not
  accounted for. 

  In its current implementation the MCMC algorithm does not allow for the density to be a
  free parameter. In principal the \ion{Ca}{15} 181.900/200.972\,\AA\ ratio should provide
  information on the density at temperatures near 4\,MK. The weaker component of this line
  pair, 181.900\,\AA, however, is not observed in these data. At the densities we expect
  ($<10^{10}$\,cm$^{-3}$) the intensity of the 181.900\,\AA\ line is less than 0.1 times
  that of the 200.972\,\AA. This implies less than 20 DN would be detected in this 60\,s
  observation. Monte Carlo simulations indicate that approximately 100 counts are needed
  to compute a line intensity. In the absence of alternatives we use the \ion{Fe}{13}
  202.836/202.044\,\AA\ ratio, which yields a density of $\log n_e = 9.45$. We will
  discuss this result later in the paper.

  The relationship between the emissivities, the differential emission measure, and the
  observed intensities is the usual expression
  \begin{equation}
    I_\lambda = \frac{1}{4\pi}\int \epsilon_\lambda(n_e,T_e)\xi(T_e)\,dT_e,
    \label{eq:ints}
  \end{equation}
  where $\epsilon_\lambda(n_e,T_e)$ is the emissivity computed with the CHIANTI 6.0.1
  atomic database assuming coronal abundances \citep{feldman1992} and the new CHIANTI
  ionization fractions \citep{dere2009}. The function $\xi(T_e)=n_e^2\,ds/dT_e$ is the
  differential emission measure distribution inferred from the intensities.  Here we also
  show the emission measure loci computed from
  \begin{equation}
    \xi_{loci}(T_e) = \frac{4\pi I_\lambda}{\epsilon_\lambda(n_e,T_e)},
  \end{equation}
  which indicates the temperature range where the various lines are sensitive. Note that
  to aid in the comparisons with the em loci we will always plot the DEM multiplied by the
  temperature bin,
  \begin{equation}
    \xi(T_e)\,dT_e,
  \end{equation}
  and we refer to this as the emission measure distribution. 

  The DEM computed from the MCMC method is shown in Figure~\ref{fig:dem}. The intensities
  computed from the DEM, which are also given in Table~\ref{table:ints1}, are generally
  consistent with the observations to within 25\%, although there are some significant
  discrepancies. We note that the Mg lines, which lie at the low end of the temperature
  range, are not fully consistent with \ion{Si}{7}. We conjecture that this is due to
  uncertainties in the abundances. Since the intensity for \ion{Si}{10} is consistent with
  the other Fe lines formed at a similar temperature, we assume that this discrepancy is
  due to the Mg abundance. To compute the emission measure we multiply the coronal Mg
  abundance by a factor of 1.7 to bring \ion{Mg}{7} 280.737\,\AA\ into agreement with
  \ion{Si}{7} 275.368\,\AA. In this analysis we have also included two lines from S, which
  is a low first ionization potential element. The consistency between the S and Fe lines
  in this DEM calculation suggests that the assumption of coronal abundances is generally
  valid, at least in a relative sense. Among the Fe lines the most significant problem is
  for \ion{Fe}{13}. This is perplexing since there are no such problems evident in the
  analysis of quiet Sun data. This issue is unresolved here, but the emission measure is
  meant to be only a crude representation of the temperature distribution in these core
  active region loops. We will consider comparisons between some simple heating models and
  the observed intensities in the next section.

  The most significant result from this calculation is the fact that the emission measure
  is strongly peaked at about $\log T_e = 6.6$ (4\,MK) and falls off sharply at both
  higher and lower temperatures. The peak emission measure is approximately
  $10^{28}$\,cm$^{-5}$ while the emission measure at $\log T_e = 5.8$, the temperature of
  formation for \ion{Si}{7}, is about a factor of 400 less.  Because of the broad
  nature of the \ion{Ca}{17} and XRT Al-thick responses, the decline in the emission
  measure at higher temperatures is less certain. The Monte Carlo simulations suggest a
  fairly broad range of possible emission measures at the higher temperatures, but the
  vast majority of them lie at the lower end of the range.

  Numerous active region emission measure distributions have been published in the
  literature
  \cite[e.g.,][]{dere1982,dere1993,brosius1996,warren2001,winebarger2011,tripathi2011}. Our
  emission measure distribution has significantly steeper slopes away from the maximum
  than most previous results. We attribute this to the fact that most previous work has
  used intensities derived from averages over large field of view. Our work, in contrast,
  has focused on a very small region that represents the conditions near the loop
  apex. \cite{tripathi2011} and \cite{winebarger2011} have also considered a small
  ``inter-moss'' region and find both shallow and steep slopes from the peak emission
  measure to lower temperatures. It is clear that a systematic study is required to
  understand the range of possible emission measure distributions near the loop apex.

  In computing the intensities in the active region core or inter-moss region we have not
  performed any background subtraction. The intensities of the high temperature lines
  outside of the core are negligible. For the cooler emission lines it is not clear which
  region to take as the background. The intensities for these lines in the dimmest areas
  of the active region are typically 10\% of the core intensity and close to the core the
  intensities are comparable. This suggests a very large range of possibilities. In all
  cases the background subtracted intensities would be lower than the values that we have
  used. This would lead to even steeper declines in the emission measure distributions at
  low temperatures. At low temperatures, the emission measure shown in
  Figure~\ref{fig:dem} is an upper bound. 

 \begin{figure*}[t!]
   \centerline{\includegraphics[clip,scale=0.5]{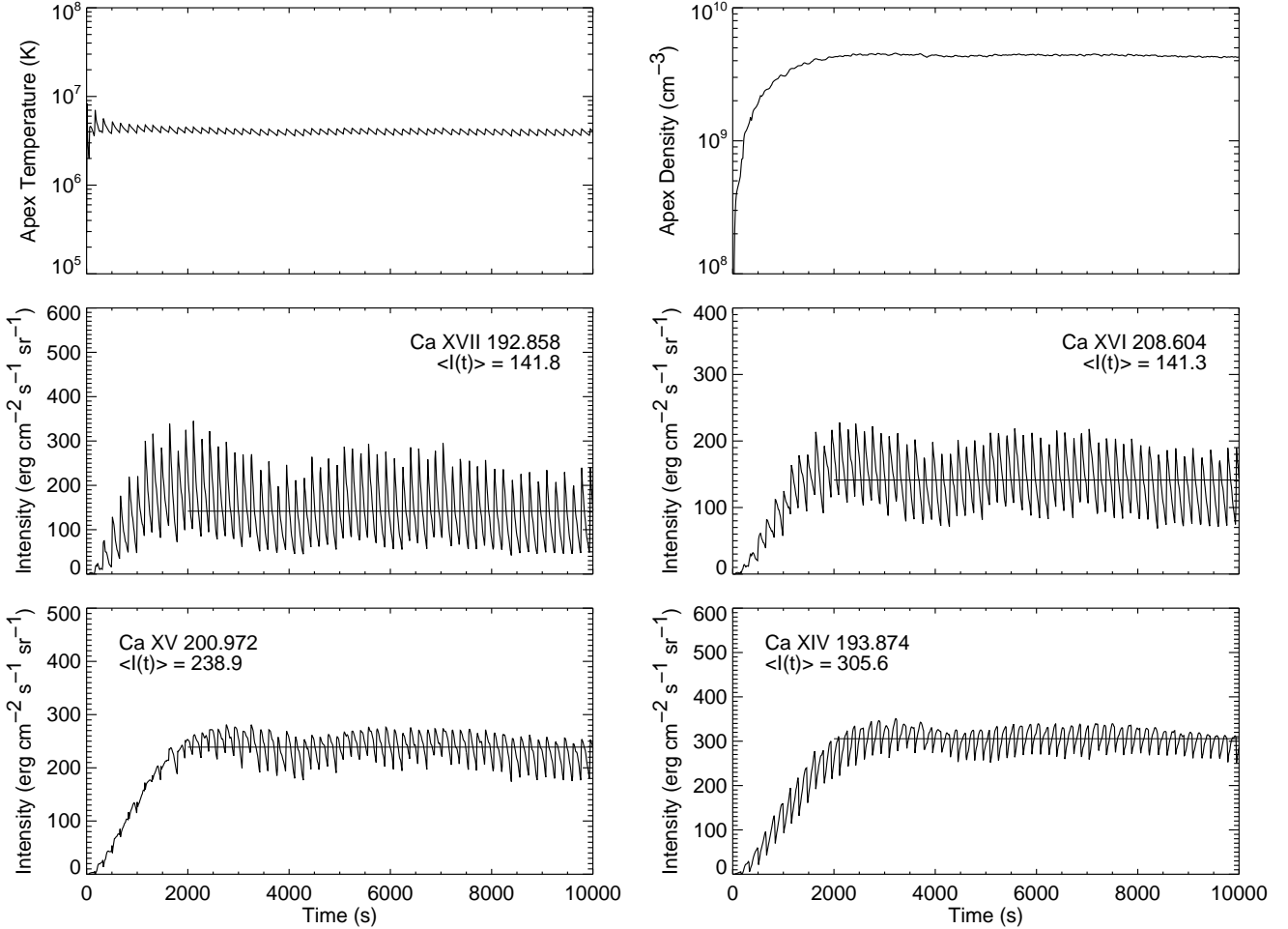}}
   \caption{An example of high-frequency heating that keeps a loop close to
     equilibrium. Shown here are the apex temperature and density (\textit{top panels}) in
     the loop and the apex intensities for four high temperature emission lines
     (\textit{bottom panels}). For the line emission the average intensity is also
     indicated. The initial atmosphere for this simulation is very tenuous (0.4 MK initial
     apex temperature) and the density converges to the asymptotic value slowly so the
     averaging is done over the final 8,000\,s of the simulation. }
   \label{fig:hf}
 \end{figure*} 

 \section{Modeling}

 Early work on the modeling of high temperature loops observed at soft X-ray wavelengths
 often focused on steady heating models \citep[e.g.,][]{kano1995,porter1995}. Such
 analysis has also yielded volumetric filling factors that were generally less than 1
 \citep{porter1995}, indicating that the observed loops were unresolved. Such observations
 could not exclude the possibility that the observed emission only appeared steady, but
 was actually made up of many unresolved loops that are evolving.  The broad temperature
 coverage of the combined EIS and XRT observations allows us to make much more detailed
 comparisons with the predictions of different heating scenarios.

 In this section we will make quantitative comparisons between the observations and
 solutions to the hydrodynamic loop equations, which describe the evolution of mass,
 momentum, and energy in a flux tube. We will consider low-frequency heating, where the
 time between heating events is long relative to the cooling time, and high-frequency
 heating, where the loop is reheated before it has time to cool. The goal is to consider
 some very simple cases that will lay the foundation for more complete studies in the
 future. The full modeling of all the observed emission in the active region
 \citep[e.g.,][]{schrijver2004,warren2007,winebarger2008,lundquist2008,winebarger2011} is
 beyond the scope of this work.

 The loop length is an important parameter in any hydrodynamic simulation. The AIA images
 suggest a distance of approximately 50\,Mm between the middle of the moss regions in this
 active region. This implies a total loop length ($L$) of approximately 75\,Mm for a
 semi-circular loop that is perpendicular to the solar surface. For simplicity we will
 consider a single loop length in the simulations. A radiative loss rate as a function of
 temperature must also be specified. The DEM calculation presented in the previous section
 showed that the observed intensities for \ion{S}{10} and \ion{S}{13} are consistent with
 coronal abundances so we will assume a radiative loss curve computed using the coronal
 abundances of \cite{feldman1992}.

 To solve the time-dependent hydrodynamic loop equations we use the NRL Solar Flux Tube
 Model (SOLFTM). We adopt many of the same parameters and assumptions that were used in
 previous simulations with this code and we refer the reader to the earlier papers for
 additional details on the numerical model (e.g., \citealt{mariska1989,warren2003}). To
 establish a characteristic time scale for the cooling of a high temperature loop with a
 total length of 75\,Mm and a coronal composition, we begin by considering an initial
 simulation of a loop in equilibrium with an apex temperature of 4\,MK, the peak
 temperature in the DEM. If the loop is allowed to cool without any additional heating the
 apex temperature reaches 0.5\,MK in approximately 1200\,s, we will use $\tau_c$ to
 represent this cooling time. 

 To model emission that persists we need to specify the volumetric heating rate as a
 function of time. We consider a series of step-function heating events that have a
 magnitude $\epsilon$, a duration $\delta$, and an occurrence rate $\tau$. We will assume
 that $\delta<\tau$ for these simulations. We will also assume that the heating is uniform
 over the loop length. The mean heating rate is simply
 \begin{equation}
   \bar{\epsilon} = \frac{\epsilon\cdot\delta}{\tau}.
   \label{eq:eps}
 \end{equation}
 For a given loop length and radiative loss curve the parameter $\bar{\epsilon}$
 determines, at least approximately, the average apex temperature for the loop. For this
 work we will assume $\bar{\epsilon} = 8.31\times10^{-3}$\,erg~cm$^{-3}$~s$^{-1}$, which
 is the heating rate required to keep the loop in equilibrium with an apex temperature of
 approximately 4\,MK, and we will vary $\epsilon$, $\delta$, and $\tau$ subject to this
 constraint. 

 An estimate of the event occurrence rate was given by \cite{cargill1994} using
 \begin{equation}
   \tau \sim \frac{Q}{\Lambda A_h},
   \label{eq:tau}
 \end{equation}
 where $Q$ is the average event energy, $\Lambda$ is the radiative loss rate of the
 corona, and $A_h$ is the spatial scale for coronal loops. For $Q\sim10^{24}$\,erg,
 $A_h\sim10^{14}$\,cm$^2$, and $\Lambda\sim10^7$\,erg~cm$^{-2}$~s$^{-1}$ the time scale is
 $\tau\sim1000$\,s. Since these parameters are simply estimates there is a considerable
 range of possible values for $\tau$. With the application of the appropriate geometrical
 factors Equation~\ref{eq:eps} can be recast in a form similar to
 Equation~\ref{eq:tau}. Our preference is to use the volumetric heating rate since it is
 the input to the hydrodynamic loop equations.

 For high-frequency heating ($\tau\ll\tau_c$) we expect the loop to be close to
 equilibrium.  To illustrate an example of such a loop we consider the parameters
 $\tau=150$\,s, $\delta=13$\,s, and $\epsilon =
 9.28\times10^{-2}$\,erg~cm$^{-3}$~s$^{-1}$. To simplify the discussion we average the
 solution over the top 50\% of the loop length at each time step to determine the apex
 density and temperature as a function of time. We then compute the time-dependent
 intensity of each emission line of interest using
 \begin{equation}
    I_\lambda = \frac{1}{4\pi} \epsilon_\lambda(n_e,T_e)n_e^2\,ds.
 \end{equation}
 We infer the path length ($ds$) from the observed \ion{Ca}{15} 200.972\,\AA\ intensity
 and use this value in calculating the intensities for all of the other lines. For this
 example the path length is 12.3\,Mm. Finally, to approximate the superposition of many
 sub-resolution strands in various stages of heating and cooling we compute the
 time-averaged intensity \citep[e.g.,][]{patsourakos2006,warren2007}. The resulting apex
 temperatures and densities, are shown in Figure~\ref{fig:hf} along with the intensities
 for the high temperature Ca lines. All of the simulated and observed intensities are
 given in Table~\ref{table:ints2}.

 \begin{figure*}[t!]
   \centerline{\includegraphics[clip,scale=0.5]{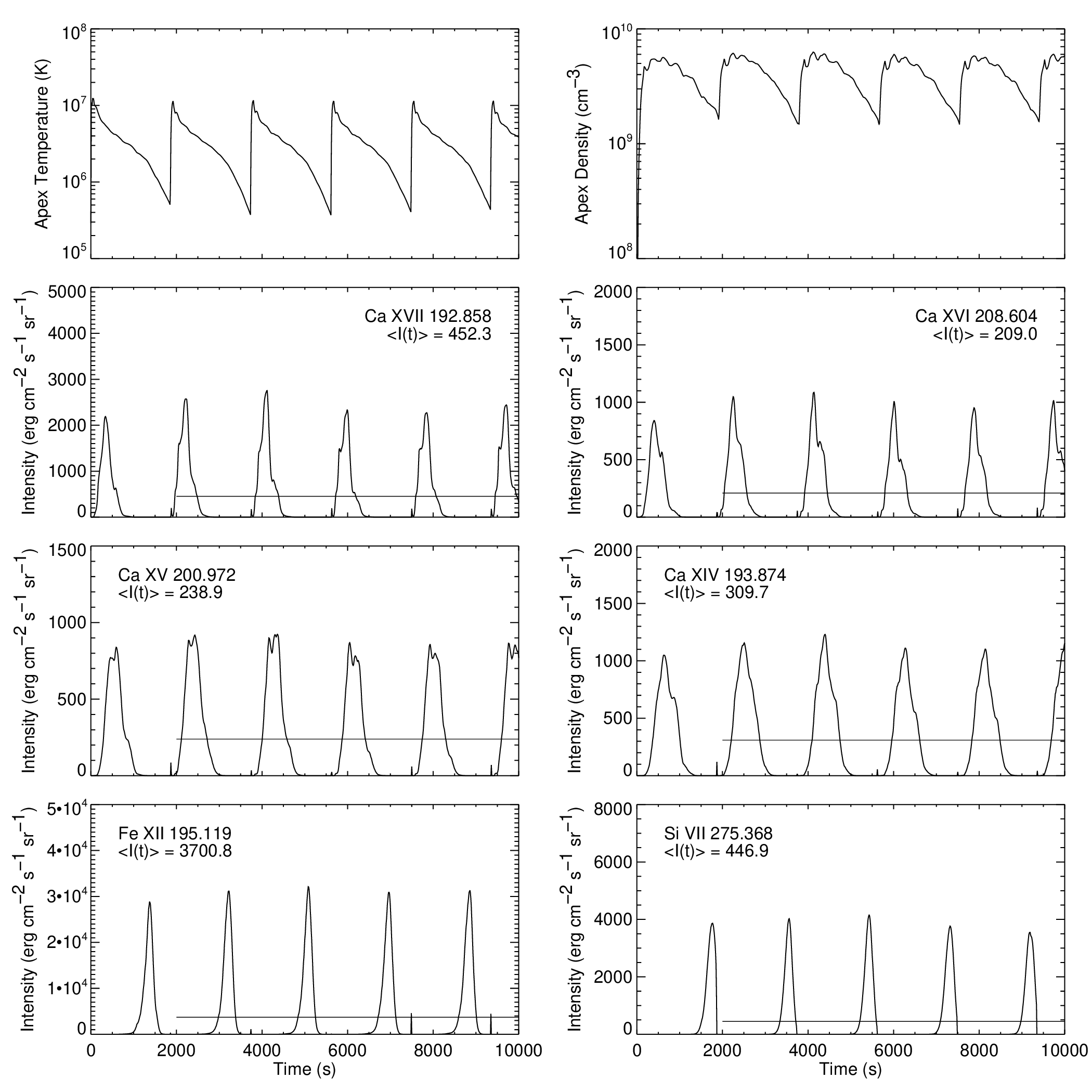}}
   \caption{An example of low-frequency heating where the long time between heating events
     allows the loop to evolve over a wide range of temperatures. The top panels show the
     temperature and density averaged over the loop apex. The bottom panels show the
     evolution of the intensities in selected emission lines. The time averaged
     intensities, again taken over the final 8,000\,s of the simulation are also
     indicated. The large intensities calculated for \ion{Fe}{12} 195.119\,\AA\ and
     \ion{Si}{7} 275.368\,\AA\ are not consistent with the observations.}
   \label{fig:lf}
 \end{figure*} 

\begin{deluxetable}{lrrrrrcrrcrr}
\tablewidth{0in}
\tabletypesize{\small}
\tablecaption{Observed and Simulated Active Region Core Intensities\tablenotemark{a}}
\tablehead{
  \multicolumn{4}{c}{}                   & 
  \multicolumn{2}{c}{$\tau=150$\,s}      &
  \multicolumn{1}{c}{}                   &
  \multicolumn{2}{c}{$\tau=1800$\,s}     &
  \multicolumn{1}{c}{}                   &
  \multicolumn{2}{c}{Mixed}              \\
[.3ex]\cline{5-6}\cline{8-9}\cline{11-12}\\[-1.6ex]
  \multicolumn{1}{c}{Line}          &
  \multicolumn{1}{c}{$T_{max}$}     &
  \multicolumn{1}{c}{$I_{obs}$}     &
  \multicolumn{1}{c}{$\sigma_{I}$}  &
  \multicolumn{1}{c}{$I_{model}$}   &
  \multicolumn{1}{c}{$R$}           & 
  \multicolumn{1}{c}{}              & 
  \multicolumn{1}{c}{$I_{model}$}   & 
  \multicolumn{1}{c}{$R$}           &
  \multicolumn{1}{c}{}              & 
  \multicolumn{1}{c}{$I_{model}$}   & 
  \multicolumn{1}{c}{$R$}
}
\startdata
\ion{Mg}{5} 276.579    &      5.45 &      16.5 &       3.8 &       0.0 &       $>$10 &  &      57.5 &       0.3 &  &       5.8 &       2.9 \\ 
\ion{Mg}{6} 270.394    &      5.65 &      35.9 &       7.9 &       0.0 &       $>$10 &  &     200.5 &       0.2 &  &      20.1 &       1.8 \\ 
\ion{Mg}{7} 280.737    &      5.80 &      32.7 &       7.4 &       0.0 &       $>$10 &  &     213.7 &       0.2 &  &      21.4 &       1.5 \\ 
\ion{Si}{7} 275.368    &      5.80 &      47.0 &      10.4 &       0.0 &       $>$10 &  &     446.9 &       0.1 &  &      44.7 &       1.1 \\ 
\ion{Fe}{9} 197.862    &      5.85 &      40.0 &       8.8 &       0.0 &       $>$10 &  &     302.2 &       0.1 &  &      30.2 &       1.3 \\ 
\ion{Fe}{9} 188.497    &      5.85 &      72.4 &      16.0 &       0.0 &       $>$10 &  &     509.1 &       0.1 &  &      50.9 &       1.4 \\ 
\ion{Fe}{10} 184.536   &      6.05 &     280.4 &      61.8 &       0.0 &       $>$10 &  &    1046.8 &       0.3 &  &     104.7 &       2.7 \\ 
\ion{Si}{10} 258.375   &      6.15 &     294.0 &      64.7 &       4.4 &       $>$10 &  &     958.3 &       0.3 &  &      99.8 &       2.9 \\ 
\ion{Fe}{11} 188.216   &      6.15 &     578.2 &     127.2 &       0.1 &       $>$10 &  &    1754.6 &       0.3 &  &     175.6 &       3.3 \\ 
\ion{Fe}{11} 180.401   &      6.15 &     926.1 &     204.2 &       0.1 &       $>$10 &  &    3507.1 &       0.3 &  &     350.8 &       2.6 \\ 
\ion{S}{10} 264.233    &      6.15 &      71.8 &      15.9 &       0.6 &       $>$10 &  &     205.0 &       0.3 &  &      21.0 &       3.4 \\ 
\ion{Fe}{12} 195.119   &      6.20 &    1475.4 &     324.6 &       2.5 &       $>$10 &  &    3700.8 &       0.4 &  &     372.3 &       4.0 \\ 
\ion{Fe}{12} 192.394   &      6.20 &     437.8 &      96.3 &       0.8 &       $>$10 &  &    1187.5 &       0.4 &  &     119.5 &       3.7 \\ 
\ion{Fe}{13} 202.044   &      6.25 &    1248.3 &     274.7 &      10.2 &       $>$10 &  &    1465.2 &       0.9 &  &     155.7 &       8.0 \\ 
\ion{Fe}{13} 203.826   &      6.25 &    2533.9 &     557.6 &      23.4 &       $>$10 &  &    3446.9 &       0.7 &  &     365.8 &       6.9 \\ 
\ion{Fe}{14} 270.519   &      6.30 &     515.0 &     113.3 &      84.6 &       6.1 &  &    1343.2 &       0.4 &  &     210.5 &       2.4 \\ 
\ion{Fe}{14} 264.787   &      6.30 &    1026.9 &     226.0 &     177.8 &       5.8 &  &    2829.7 &       0.4 &  &     443.0 &       2.3 \\ 
\ion{Fe}{15} 284.160   &      6.35 &   10334.0 &    2273.6 &    6823.7 &       1.5 &  &   23716.4 &       0.4 &  &    8513.0 &       1.2 \\ 
\ion{S}{13} 256.686    &      6.40 &     854.7 &     188.1 &     603.4 &       1.4 &  &    1567.7 &       0.6 &  &     699.8 &       1.2 \\ 
\ion{Fe}{16} 262.984   &      6.45 &    1157.6 &     254.7 &    1078.9 &       1.1 &  &    1741.8 &       0.7 &  &    1145.2 &       1.0 \\ 
\ion{Ca}{14} 193.874   &      6.55 &     311.9 &      68.6 &     305.6 &       1.0 &  &     309.7 &       1.0 &  &     306.0 &       1.0 \\ 
\ion{Ca}{15} 200.972   &      6.65 &     238.9 &      52.6 &     238.9 &       1.0 &  &     238.9 &       1.0 &  &     238.9 &       1.0 \\ 
\ion{Ca}{16} 208.604   &      6.70 &     122.0 &      28.3 &     141.3 &       0.9 &  &     209.0 &       0.6 &  &     148.1 &       0.8 \\ 
\ion{Ca}{17} 192.858   &      6.75 &     146.5 &      32.3 &     141.8 &       1.0 &  &     452.3 &       0.3 &  &     172.9 &       0.8 \\ 
XRT Open/Al-thick      &      7.10 &       4.5 &       0.9 &       6.3 &       0.7 &  &      14.9 &       0.3 &  &       7.2 &       0.6
\enddata
\tablenotetext{a}{All intensities are in units of erg cm$^{-2}$ s$^{-1}$ sr$^{-1}$. The
  parameter $R$ is the ratio of the observed to modeled intensities and $\tau$ is time
  between heating events. Mixed refers to a combination of 90\% high frequency and 10\%
  low frequency intensities.}
\label{table:ints2}
\end{deluxetable}

 For this set of parameters the high temperature emission lines (above $\log T = 6.45$)
 are generally well produced by the model. Since heating events occur very frequently the
 the temperatures and densities in the loop are never too far from their average values
 ($\log T = 6.60$ and $\log n_e = 9.64$), and the loop never cools. These simulation
 results are from a family of simulations where the heating duration $\delta$ and heating
 rate $\epsilon$ were varied. For larger values of delta the departures from the average
 temperature is smaller and the intensity in \ion{Ca}{17} becomes much smaller than what
 is observed. Similarly, for smaller values of delta the intensities in \ion{Ca}{17} are
 larger and also inconsistent with what is observed.

 The intensities for the lower temperature emission are not reproduced by this
 model. Since the loop never cools, the modeled intensities at low temperatures are close
 to zero. One possible interpretation is that the relatively weak cool emission is
 unrelated to the high temperature active region loops and comes from either very
 low-lying loops or very long loops that extend over the active region.

 For $\tau\gtrsim\tau_c$ the loop will evolve over a large range of temperatures and will
 be far from equilibrium. To illustrate an example of the low-frequency heating that is
 usually assumed in nano-flare modeling we consider the parameters $\tau=1800$\,s,
 $\delta=67$\,s, and $\epsilon = 2.22\times10^{-1}$\,erg~cm$^{-1}$~s$^{-1}$.  The apex
 temperatures and densities from this simulation are shown in Figure~\ref{fig:lf}. These
 parameters give approximately the same mean temperature and density ($\log T = 6.54$ and
 $\log n_e = 9.62$) as in the previous case. In addition to the high temperature Ca lines
 we also show the evolution of \ion{Fe}{12} 195.119\,\AA\ and \ion{Si}{7} 275.368\,\AA\ in
 Figure~\ref{fig:lf}. The path length inferred from the \ion{Ca}{15} intensity is
 25.2\,Mm. All of the simulated and observed intensities are given in
 Table~\ref{table:ints2}.

 In this case the long time between heating events allows the loop to cool and there is
 significant emission for lines formed at low temperatures. These intensities, however,
 are much larger than what is observed. For \ion{Si}{7} 275.368\,\AA, for example, the
 modeled intensity is almost a factor of 10 times too large. The difficulty for the
 low-frequency heating scenario is the slow draining of the density from the
 loop. Previous work has suggested an $n_e\sim\sqrt{T_e}$ relationship
 (\citealt{jakimiec1992}; see also the more recent work by \citealt{bradshaw2010} which
 refines this scaling law). This leads to relatively large densities and thus large
 intensities at low temperatures.

 The low-frequency heating case also has difficulties with the high temperature
 emission. The modeled intensities for \ion{Ca}{17} 192.858\,\AA\ and XRT Al-thick are
 larger than what is observed. This simulation is from a family of simulations with
 different values for $\delta$ and $\epsilon$. For the cases we considered
 ($\delta<100$\,s) the intensities for this emission was always about a factor of 3
 greater than what is observed. This seems to be an inevitable consequence of low
 frequency heating. To produce relatively high densities with short bursts of heating the
 magnitude must be large, which leads to high temperatures and large \ion{Ca}{17}
 192.858\,\AA\ and XRT Al-thick intensities. 

\begin{figure}[t!]
   \centerline{\includegraphics[clip,scale=1.00]{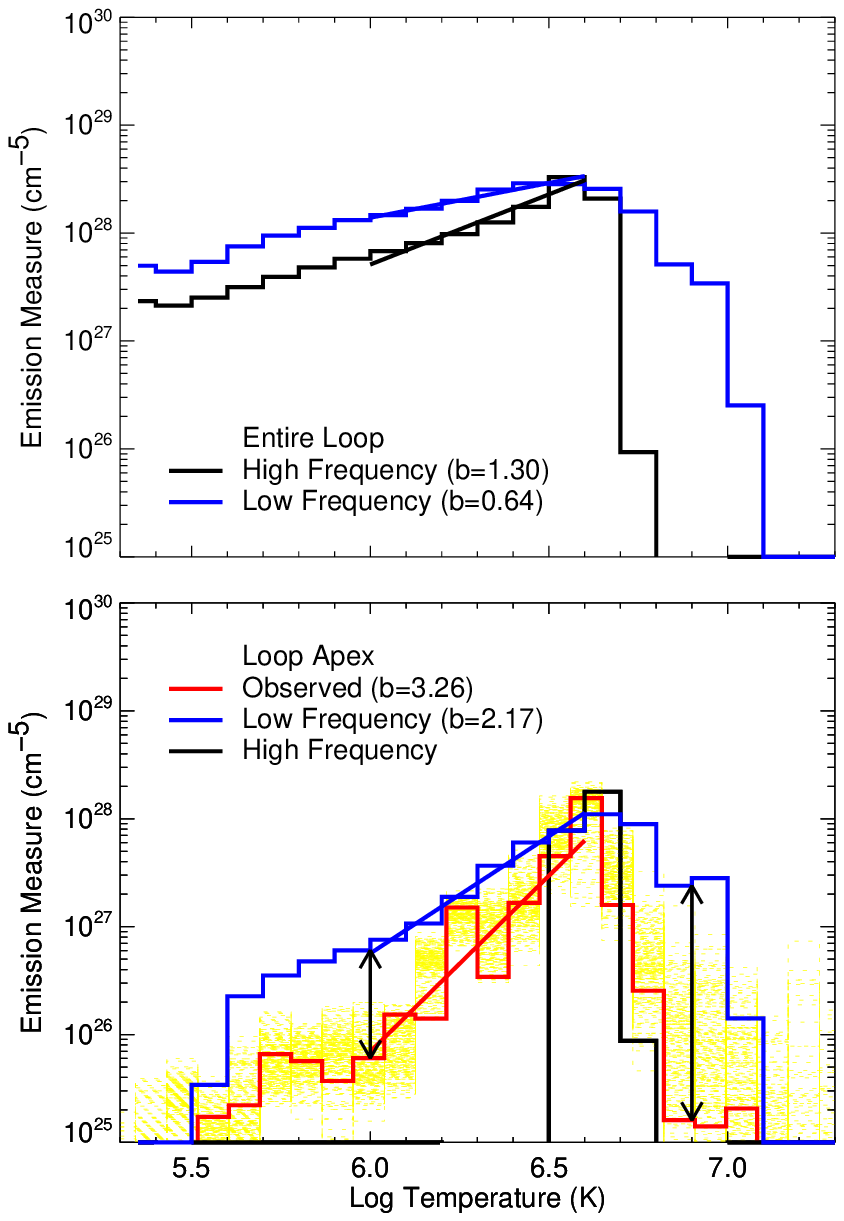}}
   \caption{The emission measure distributions derived from the high- and low-frequency
     heating simulations. The distribution for both the entire loop (\textit{top panel})
     and loop apex (\textit{bottom panel}) are shown. For comparison, our observed
     emission measure distribution is also shown in the bottom panel. The arrows indicate
     the differences between the observation and the low-frequency model at $\log T = 6.0$
     and 6.9. The power law indexes ($\textit{EM}\sim T^b$) are indicated for several of
     the emission measure distributions.}
   \label{fig:dem2}
 \end{figure}

 As mentioned previously, the power law index that we measure in the emission measure
 distribution between the peak near $\log T = 6.6$ and lower temperatures is steeper than
 most previous measurements and steeper than is expected from simple heating models, which
 can have a emission measure that scales like $\sim T^{3/2}$
 \cite[e.g.,][]{jordan1976}. This is likely a result of averaging over a small region near
 the loop apexes in the observations. To demonstrate this we have calculated the emission
 measure distributions, essentially histograms of $n_e^2ds$, from both the high- and
 low-frequency heating simulations. These distributions, which are shown in
 Figure~\ref{fig:dem2}, clearly illustrate the effect of averaging. The power-law index
 ($b$ in $\textrm{EM}\sim T^b$) for the entire loop is smaller ($b\lesssim1.5$) than for
 the loop apex ($b\gtrsim2$).
 
 One objection that could be raised regarding the emission measure distribution is that
 the box used to compute the average intensities is small and these values are not
 representative. The contrast between the moss and inter-moss regions at low temperatures,
 however, is small. For \ion{Si}{7} 275.368\,\AA, for example, the intensities only
 increase by a factor of about 2 in the moss. It is clear that the large discrepancy
 between the intensities computed from the low-frequency heating scenario would not be
 changed significantly by modifying the region used for the averaging.

\begin{figure}[t!]
   \centerline{\includegraphics[clip,scale=1.0]{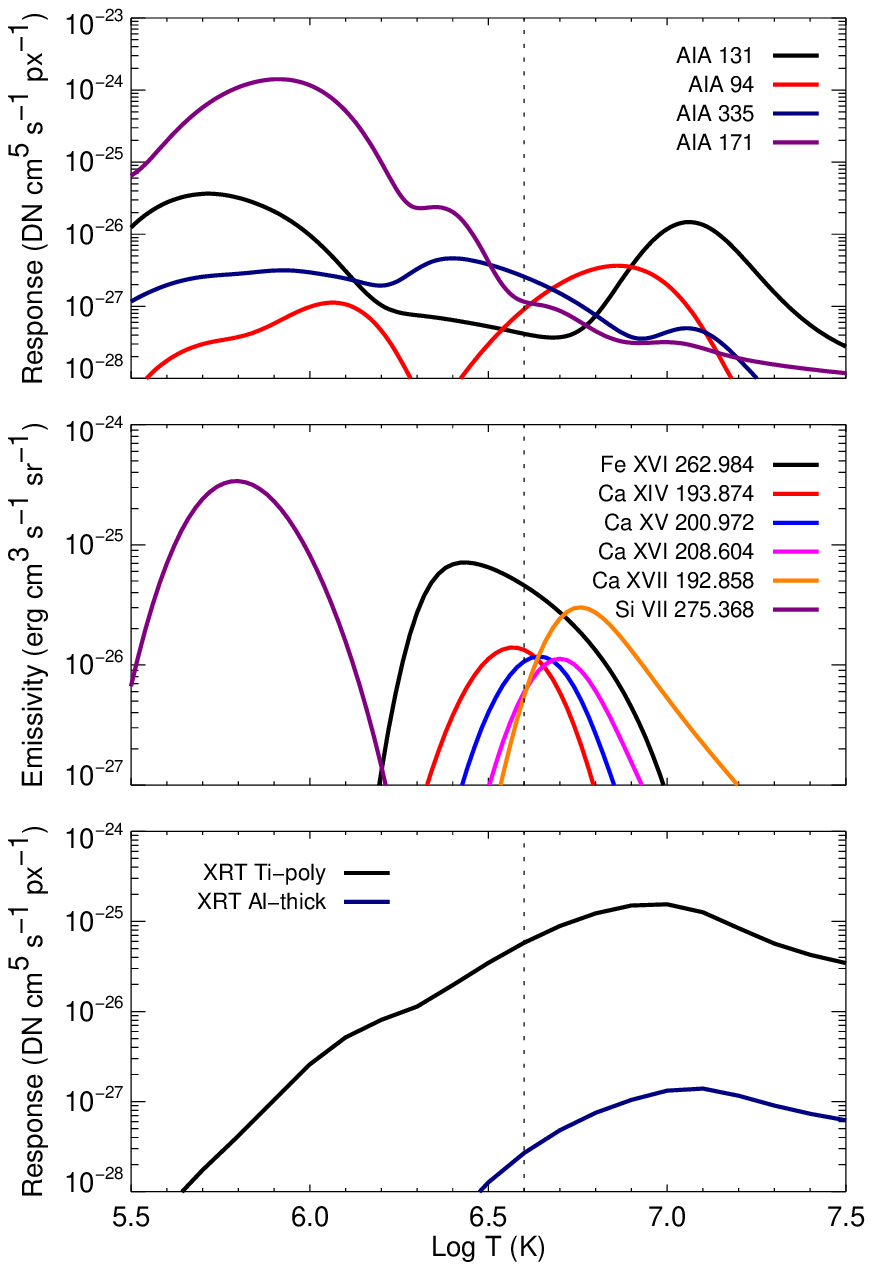}}
   \caption{Response curves for AIA, EIS, and XRT illustrating the temperature sensitivity
   of the various images and emission lines. The dotted vertical line indicates the peak
   temperature in the emission measure distribution. }
   \label{fig:resp}
 \end{figure}

 Since the rasters for the lowest temperature lines are relatively noisy, it is tempting
 to think that these lines are simply too faint to measure reliably. As we illustrate in
 Figure~\ref{fig:resp}, however, the emissivity for \ion{Si}{7} 275.368\,\AA, which is one
 of the primary EIS lines at low temperatures is actually higher than the emissivity for
 the high temperature \ion{Fe}{16} 262.984\,\AA\ line, which is easily observed in the
 core of the active region. The effective area for EIS at these wavelengths is also very
 similar.  Absent some fundamental flaw in the hydrodynamic models, it appears that any
 hot plasma evident in the \ion{Fe}{16} 262.984\,\AA\ line would be easily observed in
 \ion{Si}{7} 275.368\,\AA\ as it cooled.

 The emissivities for the hot lines shown in Figure~\ref{fig:resp} suggest similar
 inconsistencies between the observations and low-frequency, impulsive heating
 scenarios. The emissivity for \ion{Ca}{17} 192.858\,\AA\ is higher than that for
 \ion{Ca}{14} 193.874\,\AA, suggesting that loops that have been heated up to very high
 temperatures ($\sim10$\,MK) and are cooling would show comparable intensities in these
 lines. Instead, the data show that the \ion{Ca}{17} emission is weaker than the emission
 from \ion{Ca}{14}. Again, the observations are difficult to reconcile with traditional
 hydrodynamic models of impulsively heated loops.
 
 \section{AIA}

 As we have already seen in Figure~\ref{fig:aia}, AIA's combination of high cadence, high
 spatial resolution, and broad temperature coverage provide new ways to explore the
 temperature structure of an active region. Of particular interest to studies of high
 temperature active region loops is the 94\,\AA\ channel, which contains the \ion{Fe}{18}
 93.94\,\AA\ formed at about 7\,MK. Also of interest is the 131\,\AA\ channel, which
 contains \ion{Fe}{20}, \ion{Fe}{21}, and \ion{Fe}{22} lines. Another channel not
 previously flown is the 335\,\AA\ channel, which images \ion{Fe}{16} 335.4\,\AA. In
 Figure~\ref{fig:resp} we show the isothermal temperature response curves for the AIA 171,
 335, 94, and 131\,\AA\ channels along with the emissivities for the EIS high temperature
 lines (\ion{Fe}{16} and \ion{Ca}{14}--\ion{Ca}{17}), and the two XRT channels considered
 in this paper.

 \begin{figure*}[t!]
   \centerline{%
     \includegraphics[clip,scale=1.0]{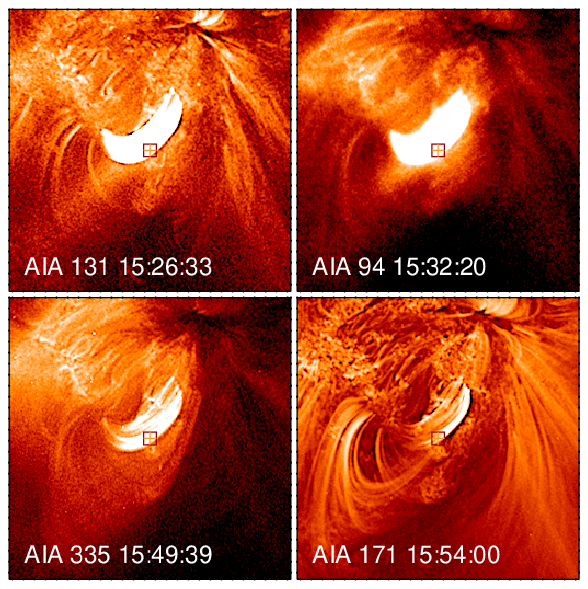}
     \includegraphics[clip,scale=1.0]{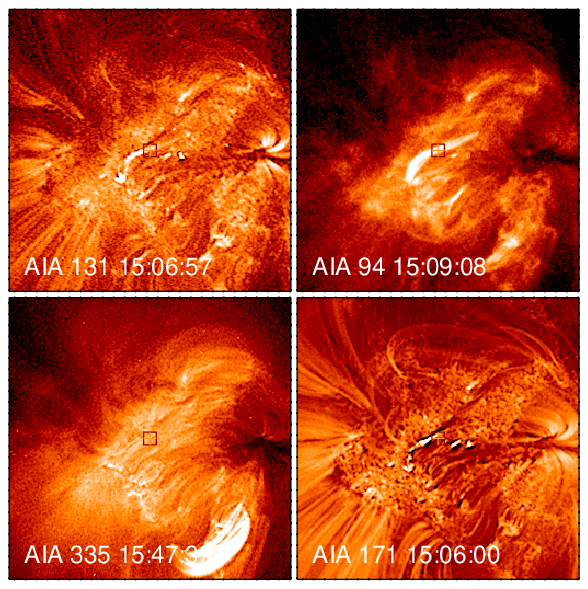}
     \includegraphics[clip,scale=1.0]{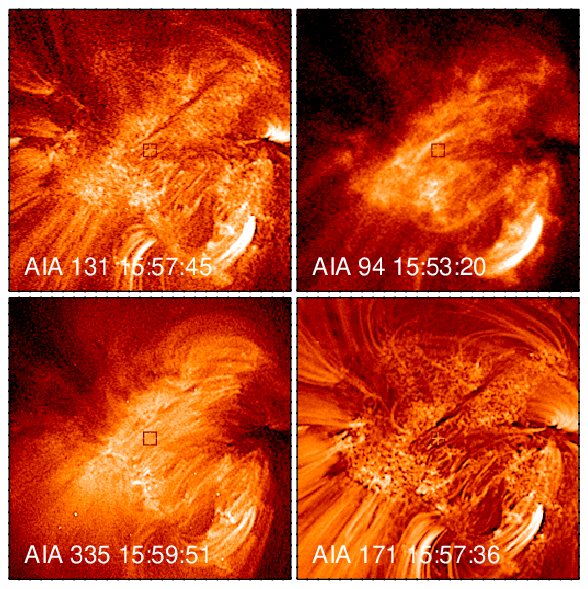}
     }
     \centerline{%
     \includegraphics[clip,scale=1.0]{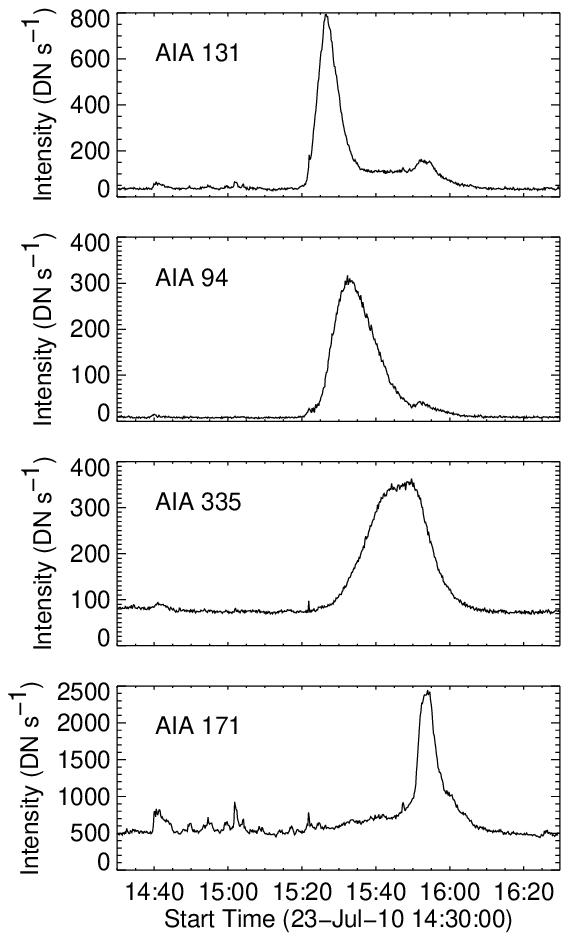}
     \includegraphics[clip,scale=1.0]{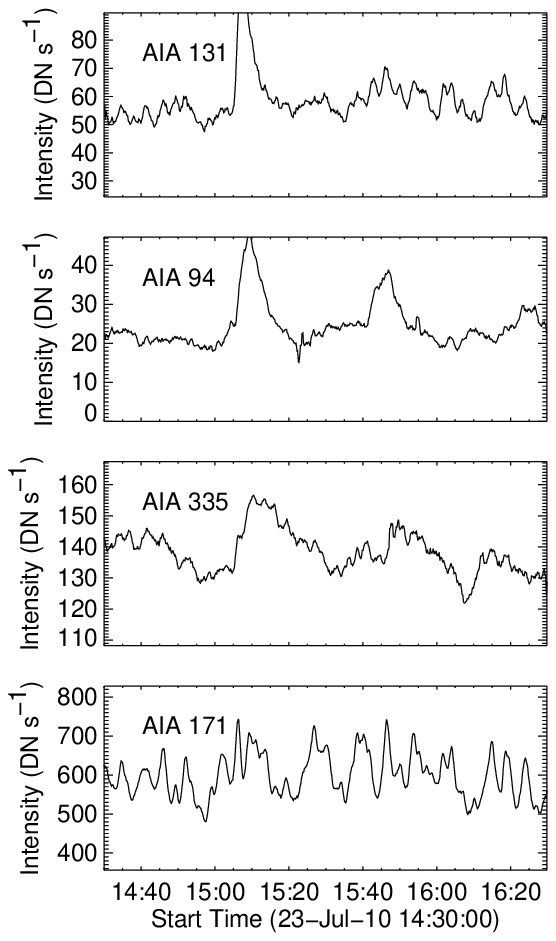}
     \includegraphics[clip,scale=1.0]{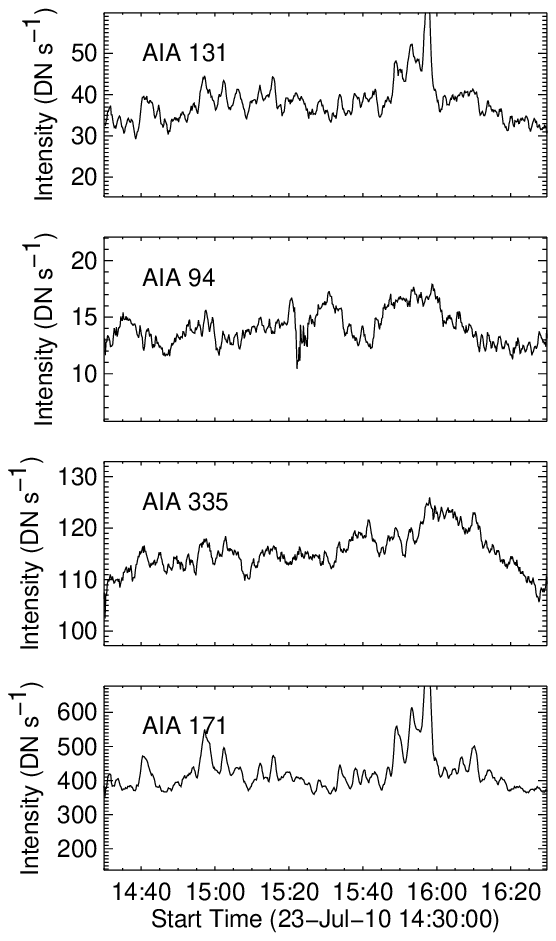}
     }
     \caption{Three representative light curves from the core of the active region. The
       top panels show the image data from the peak of the light curve. The bottom panels
       show the intensities as a function of time, the intensities have been smoothed with
       60\,s running averages. The first column shows the evolution of a small flare, with
       a clear progression from high temperatures (\ion{Fe}{20}--\ion{Fe}{23}) to low
       temperatures (\ion{Fe}{9}). The middle column shows a similar behavior for a
       microflare. The final column shows the time history of a relatively quiescent
       pixel in the active region core.}
   \label{fig:lc}
 \end{figure*}

 The AIA 94\,\AA\ channel is of particular interest because it combines high spatial and
 temporal resolution with a relatively narrow temperature response. The EIS emission lines
 are formed at somewhat lower temperatures and, at least for these observations, have no
 temporal resolution. XRT observes these temperatures at high cadence, but has a very
 broad temperature response. Since the cross-calibration of the AIA instrument is just
 beginning at this point, we're interested only in some very basic questions, such as what
 is the morphology of the emission in this channel and how does it evolve with time? 

 To provide some context for interpreting the AIA 94\,\AA\ and 131\,\AA\ channels we have
 used the simulations results for the high and low frequency heating cases shown in
 Figures~\ref{fig:hf} and \ref{fig:lf} to compute the expected count rates in these
 channels. As one would expect, we find that the amount of high temperature emission
 observed in these channels is sensitive to the heating time scale. For the low-frequency
 heating case, which reaches temperatures close to 10\,MK, observable emission
 ($\sim50$--100~DN~s$^{-1}$) is predicted for both channels. The high-frequency case shows
 some signal in the 94\,\AA\ channel but relatively little in 131\,\AA. These comparisons
 suggest that these channels will provide important information on the heating time
 scale. Quantitative comparisons, however, will require the AIA calibration to be more
 fully understood. Also, at these high temperatures the potential for departures from
 ionization equilibrium are much greater. 

 It is clear from the movie of the 94\,\AA\ channel, which is available in the electronic
 version of the manuscript, that there is significant emission from \ion{Fe}{18} in the
 core of the active region. This channel, however, also contains contributions from lines
 formed at lower temperatures. Comparisons with the 171\,\AA\ channel allow us to identify
 the high temperature, \ion{Fe}{18} loops. Such comparisons suggest that the bulk of the
 \ion{Fe}{18} emission lies at the inner edge of the moss on loops that connect directly
 across the neutral line.
 
 To address the question of temporal evolution we have taken the co-aligned and de-rotated
 data cubes used to make the movies and computed light curves for various points in the
 active region. Light curves for three points are shown in Figure~\ref{fig:lc}. One of the
 points illustrates the evolution of a small flare. Here the light curves show the
 progression from the high temperature flare emission of \ion{Fe}{20}--\ion{Fe}{23}, to
 the hot \ion{Fe}{18} and \ion{Fe}{16}, to the relatively cool \ion{Fe}{9}. This
 qualitative agreement between the observations and our expectations for a flare light
 curve suggests that the emission lines contributing to each channel are properly
 identified. Other small ``microflaring'' events show a similar progression, except that
 the cooling to the lowest temperatures is difficult to identify (middle panel). Finally,
 the majority of the pixels in the core of the active region do not show coherent
 behavior. In these pixels the intensities measured in the 131 and 171\,\AA\ channels are
 well correlated, suggesting an absence of plasma at flare temperatures. For these points
 the 94 and 335\,\AA\ channels do not show any obvious correlation. There is no general
 tendency for brightenings in the 94\,\AA\ channel to be followed by brightenings in
 335\,\AA. This is potentially consistent with the high-frequency heating scenario, but
 will require more quantitative analysis to demonstrate conclusively.

 \section{Discussion}

 We have presented the detailed analysis of high temperature active region core loops
 using observations from \textit{Hinode} and \textit{SDO}. The primary result is the
 calculation of the emission measure, which shows a sharp peak at about 4\,MK and a steep
 decline at both higher and lower temperatures. These observations provide a difficult
 challenge to the low-frequency heating scenarios that have traditionally been associated
 with nanoflare heating models of the corona. As we've shown with some very simple
 hydrodynamic simulations, such models predict significant emission at lower temperatures,
 but the observed emission measure below 1\,MK is generally small in the core of an active
 region near the loop apexes. A high-frequency heating scenario, in contrast, agrees with
 the observed intensities of the high temperature emission lines, is consistent with the
 relatively steady emission observed in XRT, and with the absence of cooling loops in the
 moss in the cool AIA channels (e.g., 171\,\AA).

 Our analysis provides compelling evidence that there is a significant population of loops
 in the core of a solar active region that are heated on very short time scales, much
 shorter than a typical cooling time for this combination of density ($\log n_e\sim 9.7$),
 temperature ($\log T_e\sim 6.6$), and loop length ($L\sim 75$\,Mm). We suggest that this
 combination of parameters is a useful set for theorists to consider in developing
 physical models of high temperature coronal loops. The challenge is to identify a
 physical mechanism that not only reproduces these conditions, but is also frequent enough
 to keep the loop at high temperatures for long periods of time.

 One limitation of the high-frequency heating scenario, however, is that it does not
 reproduce the emission that is observed at low temperatures. In the emission measure
 distributions shown in Figure~\ref{fig:dem2}, for example, it is clear that the result
 from the high-frequency simulation is close to observed distribution at high temperatures
 (above $\log T = 6.5$), but significantly under-predicts the distribution at low
 temperatures. One possibility is that there is a range of heating time scales at work in
 the solar atmosphere. \cite{brooks2008} and \cite{lee2010}, for example, have suggested a
 relationship between magnetic topology and the variability of the heating. Transient
 brightenings appear to be associated with magnetic complexity, while magnetically simple
 regions show relatively constant emission. Motivated by the emission measure
 distributions shown in Figure~\ref{fig:dem} we have constructed composite intensities
 with 90\% high-frequency heating and 10\% low-frequency heating. As is shown in
 Table~\ref{table:ints2}, this simple mixture provides a better match to the observations
 at both high and low temperatures. This doesn't provide a consistent description for all
 of the lines, but does suggest how a complete description of the observed emission might
 be achieved.

 We stress that the hydrodynamic simulations presented here represent a very small set of
 possible heating scenarios. Some 0D impulsive heating models have produced relatively
 steep slopes in the emission measure distribution
 \citep[e.g.,][]{cargill1994,klimchuk2008}. We have had trouble reproducing these results
 with hydrodynamic simulations and we are currently investigating this systematically.
 Other considerations, such as variations in loop length along the line of sight,
 differences in initial conditions, or accounting for non-equilibrium ionization, may
 bring low-frequency, nanoflare heating models into better agreement with the
 observations. Much more extensive analysis is required. One significant hurdle that all
 models must clear is reproducing the observed intensities both in the corona and at the
 footpoints. The next step in this analysis to couple hydrodynamic simulations to topology
 derived from magnetic field extrapolations and attempt to reproduce all of the emission
 observed in the active region.


\acknowledgments Hinode is a Japanese mission developed and launched by ISAS/JAXA, with
NAOJ as domestic partner and NASA and STFC (UK) as international partners. It is operated
by these agencies in co-operation with ESA and NSC (Norway). The authors would like to
thank the referee for many valuable comments on the manuscript.


\end{document}